# GWAT: The Geneva Affective Picture Database WordNet Annotation Tool

Marko Horvat[1], Dujo Duvnjak[2] and Davor Jug[1]

[1]Polytechnic of Zagreb, Department of Computer Science and Information Technology
[2]University of Zagreb, Faculty of Electrical Engineering and Computing, Department of Telecommunications
E-mail: marko.horvat@tvz.hr

**Abstract** - The Geneva Affective Picture Database WordNet Annotation Tool (GWAT) is a user-friendly web application for manual annotation of pictures in Geneva Affective Picture Database (GAPED) with WordNet. The annotation tool has an intuitive interface which can be efficiently used with very little technical training. A single picture may be labeled with many synsets allowing experts to describe semantics with different levels of detail. Noun, verb, adjective and adverb synsets can be keyword-searched and attached to a specific GAPED picture with their unique identification numbers. Changes are saved automatically in the tool's relational database. The attached synsets can be reviewed, changed or deleted later. Additionally, GAPED pictures may be browsed in the tool's user interface using simple commands where previously attached WordNet synsets are displayed alongside the pictures. Stored annotations can be exported from the tool's database to different data formats and used in 3rd party applications if needed. Since GAPED does not define keywords of individual pictures but only a general category of picture groups, GWAT represents a significant improvement towards development of comprehensive picture semantics. The tool was developed with open technologies WordNet API, Apache, PHP5 and MySQL. It is freely available for scientific and non-commercial use.

## I. INTRODUCTION

Affective multimedia databases are repositories of multimedia documents with annotated semantics and emotion. They are primarily used as standardized tools for stimulating specific emotional responses in exposed subjects. Affective multimedia databases are frequently employed in psychology and neuroscience for research of subjective, psycho-physiological, behavioral and neurophysiological reactions induced by viewing affective stimuli (per example [1] [2] [3]).

The landscape of affective multimedia databases is very diverse with many different and structurally unrelated databases [4] [5]. Currently, affective multimedia databases are created by manual annotation of semantics and emotion in which volunteers record their personal impressions using standardized questionnaires [6] after being induced by stimuli. Obtaining affective annotations always includes a psychological experiment conducted with a statistically relevant group of participants. Software tools for querying affective multimedia databases and extraction of stimuli are virtually nonexistent [7]. The databases are still being mostly searched manually by experts in a lengthy, repetitive and labor intensive process [7].

Typically, affective multimedia databases have very simple structures constituting of a multimedia repository and a description file [4] [5]. The repository is a container with uniquely indexed multimedia which is referenced in the description file. The repository has a very simple implementation as a file system folder – databases must be acquired from their owners and installed on a user's computer. All multimedia metadata, emotion and semantic annotations are stored in the description file which is also located in the repository as a formatted text or comma separated values (CSV) file. The format of the description file differs between databases; no two databases share exactly the same format.

GAPED is a relatively new and large affective multimedia database with 730 pictures [8]. It has an even simpler architecture than most other databases. Pictures are stored in six separate folders each compromising one semantic category: "snakes", "spiders", "human concerns", "animal mistreatments", "neutral" and "positive". The first four categories are emotionally negative. All pictures have unique names by which they are referenced and emotionally described. However, there are absolutely no semantic annotations and objects or events in pictures are not specified. The semantic content of pictures can only be vaguely implied by their semantic category membership, and a more detailed context can only be obtained by accessing and observing each individual picture.

GWAT was developed with a motivation to improve the GAPED semantically sparse dataset and to provide an opportunity to reuse GAPED content in other domains which require a more expressive multimedia description. The tool is free for academic and nonprofit use and can be obtained by contacting the first author.

The remainder of this paper is organized as follows; Section 2 gives an overview of the tool and its system architecture. Sections 3 and 4 illustrate how the multimedia content is displayed in the user interface and how to browse through pictures, together with already existing GAPED-WordNet annotations. Section 5 provides information on how to retrieve WordNet synsets in GWAT and use them to annotate GAPED pictures. Finally, Section 6 concludes the paper and provides insight into future work.

## II. GWAT ARCHITECTURE

The NAPS Search Tool is a thin-client web application written in PHP 5.4 programming language. The application requires a web server (e.g. Apache HTTP Server) configured to run PHP 5.x, or newer versions, and MySQL 5.0 database to be preinstalled on the target computer. The tool is distributed as a single package with PHP pages, presentation files and a SQL script for MySQL database import. The distribution package also contains a help file with instructions on how to install the tool, configure the web server and create a new instance of GWAT relational database.

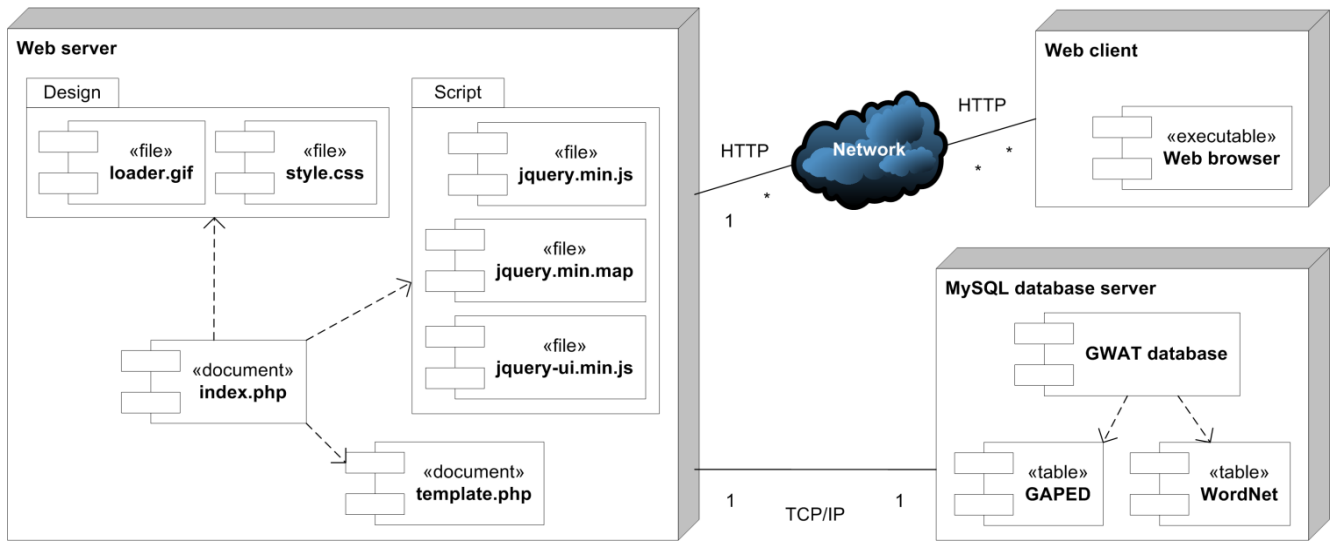

Fig. 1. GWAT's UML component and deployment diagrams.

The architecture is simple and consists of the presentation layer, the logic layer and the data layer. Cascading Style Sheet (CSS), graphics and JQuery scripts form the presentation layer, while PHP5 code files implement application logic and connect the data layer with the presentation. GAPED picture identifiers, WordNet synsets and pictures annotations created with GWAT are stored together in a single MySQL database. Pictures and affective ratings are not provided with the tool and have to be obtained from the Swiss Center for Affective Sciences[1].

WordNet semantic network is employed as a dictionary to describe semantics of GAPED pictures [9]. WordNet represents a very important component of the tool's architecture. WordNet is a large knowledge taxonomy with >100,000 individual concepts or synsets (i.e. synonym sets) connected by different semantic relationships such as IS-A and PART-OF. It is freely available and comes with an extensive software development interface. The glossary, originally in English but subsequently translated to a number of other languages, contains uniquely referenceable nouns, verbs, adjectives and adverbs. WordNet offers many benefits to description of multimedia compared to plain keywords and user defined tags. Firstly, it has very large and sufficiently expressive vocabulary. Since the glossary is controlled and standardized, picture annotations can be transferred to any other application compatible with WordNet. Finally, WordNet glossary enables users to search queries in a linguistically more natural and expressive manner as already demonstrated by existing WordNet stimuli annotating tools (per example [10] [11]).

The GWAT's UML component and deployment diagrams are shown together in Fig. 1. The main PHP page is "index.php". It uses "template.php" as a modular component for presentation of different GAPED pictures, three JQuery files for client-side functionality and CSS definition file with an animated GIF picture for design.

The program user interface has six major components:
1) the selected stimulus component for displaying a single stimulus,
2) GAPED search box,
3) GAPED navigation buttons,
4) WordNet annotations component for displaying the list of synsets describing picture semantics,
5) WordNet component for searching and attaching synsets to pictures,
6) help box with useful information for GWAT users.

The components' functionalities are described in next sections.

## III. DISPLAYING GAPED CONTENT

After starting the web tool, i.e. opening the main page "index.php", first GAPED picture A001.bmp is always automatically loaded and displayed in the selected stimulus component for displaying a single stimulus. The component is placed in the central region of the main page. If GAPED database is not installed on the server a busy JavaScript animated icon will be visible instead of the picture. All pictures can be accessed and displayed in the tool's main page together with their existing WordNet annotations. The layout of the main page is explained in the Fig. 2.

The name of the picture file with its extension is shown above and to the left of the displayed picture, and GAPED search box is on the right. User can enter name of a picture, from any of GAPED folders containing different semantic categories ("A", "H", "N", "P" "Sn" and "Sp"), to be loaded and displayed in the main page. It is not necessary to specify a specific folder because the tool will search all folders until the desired picture is found. Picture format extension is obligatory. It is necessary to specify picture names exactly – special search characters such as "*", "?" etc. are not supported in the current version of GWAT.

---
[1] http://www.affective-sciences.org/

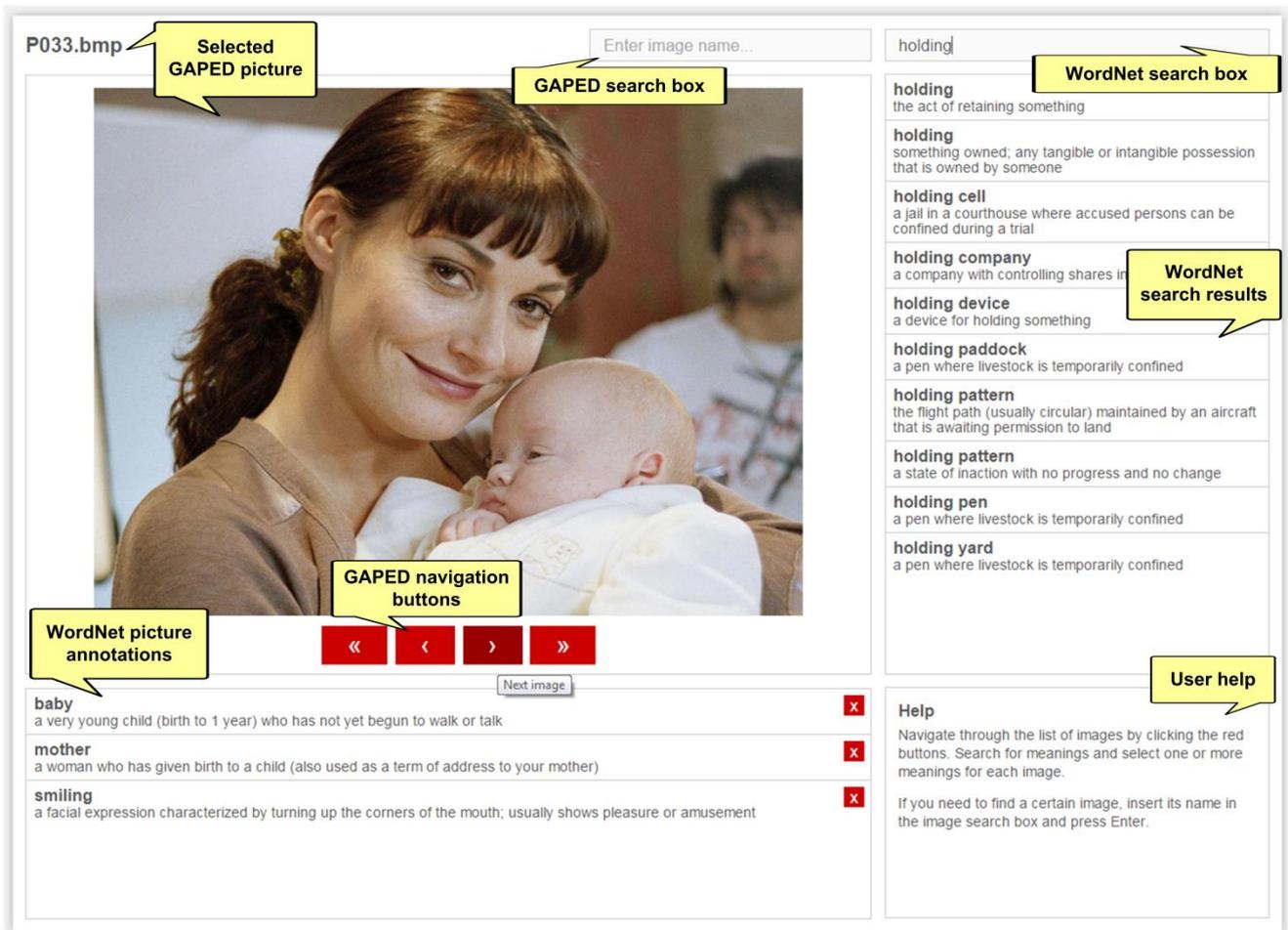

Fig. 2. The GWAT's main page.

The picture search starts only after pressing enter key and not after each letter keystroke in order to minimize resource requirements from quickly repetitive database queries. If a picture cannot be found, i.e. search query returns NULL value, a JavaScript generated (i.e. client-side) error message is shown informing the user that the particular picture cannot be found.

Below the central area are four navigation buttons (Fig. 2). The left most button ("<<") shifts the focus to the first picture in alphabetical order ("A001.bmp"). The right most button (">>") moves the focus to the last picture in the set ("Sp160.bmp"). Buttons ("<") and (">") change current display by one position towards lower position or higher position in the alphabetically sorted sequence, respectively. For example, if picture "P033.bmp" is shown (as in Fig. 2) then by pressing two times button ("<") "P032.bmp" will be displayed first followed by "P031.bmp". After pressing (">") the selection returns to "P032.bmp". The navigation buttons can also be used to circulate the view between both ends of the set: if the last picture "Sp160.bmp" is selected then with (">") the first picture "A001.bmp" will be displayed, and vice versa with ("<"). Tooltips are provided for all buttons so users can get accustomed to the tool's interface more easily.

IV. GAPED-WORDNET ANNOTATIONS

Below the navigation buttons is the WordNet annotations component with a list of WordNet synsets assigned to the currently displayed GAPED picture. The content of the component is refreshed dynamically whenever a new picture is shown in the main page. Synsets are displayed in rows with their name, full description and a large "X" button to the right (Fig. 2). By pressing the red "X" button the particular annotation is immediately removed from semantic description of the currently shown picture. This feature is implemented with JQuery JavaScript framework. There is no undo action so the synset has to be attached again if it was removed in error.

Initially the component is always empty because the GWAT database does not contain any annotations, but as − in the course of working with the tool − new semantic descriptions are generated they will be made visible in the component.

The annotations are displayed together in the component interface, sorted alphabetically and grouped based on their lexical type (noun, verb, adjective and adverb). The database is designed so that, at least in theory, there is no limit to the number of synsets that can be assigned to a single picture.

All attached synsets can be exported from the tool's database as a SQL script containing CREATE TABLE and ADD ROW statements with identifications (IDs) of GAPED pictures and WordNet synsets associated with SQL foreign key relationship. Using the script it is possible to transfer the generated annotations to other applications or 3rd party systems compatible with SQL standard and WordNet semantic network.

## V. USING WORDNET TO ANNOTATE GAPED PICTURES

WordNet search component is located on the right side of the main page. Search box is on the top of the component and a frame with search results below. The component is implemented by almost exclusively relying on JQuery JavaScript framework functionalities. The search starts on each keystroke, i.e. when browser registers keyboard press event, and depending on the user's computer performance several seconds are needed for results to be retrieved from the database. The waiting time is inversely proportional to the length of WordNet query − if only a few letters are used in search the retrieval process takes longer − and becomes progressively shorter as the query gets longer. The busy JavaScript animated icon is visible during retrieval execution indicating that the query has been accepted by the system and is being processed. With regards to the implementation the search is executed by JQuery JavaScript framework dynamically querying MySQL database, i.e. directly from the client to the server without any additional server-side PHP code.

The WordNet search results are presented as a vertical list with each row in a separate frame containing the name and description of one found synset. To improve the search component's usability synsets are sorted alphabetically and, additionally, those with the same lexical type are grouped together. If a search returns synsets of different lexical types they will be sorted in the following order: nouns first, adjectives second, verbs third and finally adverbs as the fourth group (Fig. 3). If many synsets are retrieved a vertical scroll bar will appear on the right side of the component allowing users to slide between different parts of the list.

When a synset is found with the desired semantic meaning user must click the row containing the synset with left mouse button to attach it to the currently displayed GAPED picture. This process is almost instantaneous, even on a computer with a modest performance, because the client code is short and on the server side only a single row has to be inserted in the table connecting the synset in WordNet table and the row in the table with GAPED pictures' names.

Once the synset attaching process is completed server-side, and the confirmation is received on the client, a new item is automatically generated in the WordNet annotations component for displaying the list of synsets describing picture semantics. The new item can be removed at any time by clicking on "X" button, if added in error, and a different synset can be attached instead.

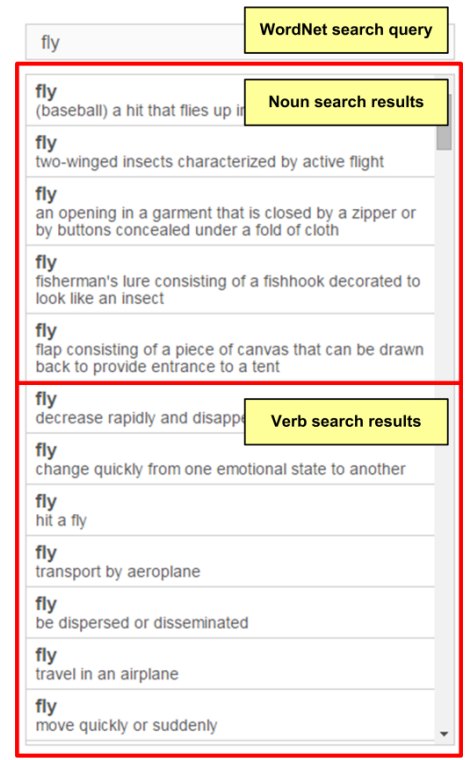

Fig. 3. WordNet search results for query "fly" in GWAT. Lexically identical synsets are grouped together.

## VI. CONCLUSION

GWAT is web-based image annotation tool for manual identification of objects and events in GAPED images and annotation with WordNet knowledge taxonomy. Generated semantic annotations are stored in a relational database where they can be modified as needed and later exported to 3rd party tools. The purpose of GWAT is to enable a more detailed semantic description of GAPED pictures with a large and standardized linguistic ontology. There are many potential applications which benefit from rich multimedia semantics. Primarily, these can be used to build multimedia search tools that take advantage of WordNet semantic similarity measures to improve quality of the retrieved dataset. The similarity measures may also be used to rank the results relative to their significance. It has been already demonstrated with other affective multimedia databases that WordNet annotations significantly improve ranking accuracy and precision in picture retrieval [4]. Secondly, GWAT generated data provides an opportunity to study the relationship between semantics and emotion in multimedia. This line of research could establish statistically relevant relationships which can then be used to automatically infer emotion of a picture or video from their semantic description. If successful, such applications could eventually lead to machine-based emotion indexing of multimedia and emotional search engines [5].
However, both of these applications are impossible with the current state of GAPED semantics and can be investigated only with tools such as GWAT.

Web image repositories and tools for semantic annotation of pictures are fairly common, but there are not many tools for construction of affective multimedia databases. Therefore, the development of GWAT is interesting as a relatively new research area in the fields of applied computing and construction of information systems.

The future work will focus on development of new GAPED search tools which use WordNet descriptions generated with GWAT. In this regard we plan to develop a web-based search tool compatible with WordNet described GAPED. Furthermore, we would like to extend GWAT so it is possible to annotate image regions and associate WordNet description to individual blobs, features or objects in pictures. GAPED pictures are high-quality and contain a wide range of semantic content so detailed annotations could be used in computer vision research for automated object recognition. Additionally, it could be expected that such more detailed annotations would also further increase the quality of picture search.